\documentclass[twocolumn]{revtex4-1}
\usepackage{graphicx}
\usepackage{color}
\newcommand{\overlap}{{\rm overlap}}

\newcommand{\vac}{|{\rm vac}\rangle}
\newcommand{\ket}[1]{\left| #1 \right\rangle}
\newcommand{\bra}[1]{\left\langle #1 \right|}
\newcommand{\braket}[2]{\left\langle #1 | #2 \right\rangle}

\newcommand{\proj}[1]{| #1\rangle\!\langle #1 |}
\newcommand{\Tr}{\mathrm{Tr}}

\newcommand{\eea}{\end{eqnarray}}
\newcommand{\bea}{\begin{eqnarray}}
\newcommand{\ee}{\end{equation}}
\newcommand{\be}{\begin{equation}}

\newcommand{\ido}{\int_{0}^\infty\!d\omega\,}
\newcommand{\tF}{\tilde{F}}

\begin{document}
\title{Detecting two photons with one molecule}

\author{Saumya Biswas and S.J. van Enk}

\affiliation{Department of Physics and
Oregon Center for Optical, Molecular \& Quantum Sciences\\
University of Oregon, Eugene, OR 97403}

\begin{abstract}
We apply input-output theory with quantum pulses [AH Kiilerich, K M\o lmer, 
Phys. Rev. Lett. {\bf 123}, 123604 (2019)] to a model of a new type of two-photon detector consisting of one molecule that can detect two photons arriving sequentially in time. 
The underlying process is distinct from the usual two-photon absorption process where two photons arriving simultaneously and with frequencies adding up to the resonance frequency are absorbed by a single molecule in one quantum jump.
Our detector model includes a Hamiltonian description of the amplification process necessary to convert the microscopic change in the single molecule to a macroscopic signal. 
\end{abstract}
\maketitle
\section{Introduction}

There are two standard ways of detecting two photons in a photon-number resolved (PNR) manner: (i) an inherent PNR detector produces a different signal depending on whether one or two photons were absorbed by the detector, (ii) multiplexed PNR detection \cite{fitch2003} exploits multiple single-photon detectors, and the signal consists of either one or two such detectors ``clicking.''  
An inherent PNR detector may, for example, be sensitive to the total energy deposited by the photons \cite{rosenberg2005}.  A second type of detector sensitive to two photons makes use of a process called ``two-photon absorption'' (TPA) in which one molecule can absorb two photons that arrive simultaneously and whose frequencies add up to the resonance frequency. This effect was discovered by G\"oppert-Mayer in 1931  \cite{goppert1931,goppert2009}, goes through a virtual intermediate state,  and has become an item of modern interest since the realization that this TPA process is sensitive to time-frequency entanglement between the two incoming photons \cite{fei1997,raymer2021,tabakaev2021}. 

For a biological example of multiplexing, we may consider the human eye. There are about $10^8$ rods, each of which is sensitive to single photons in that they can absorb one photon at a time \cite{okawa2007}.  Interestingly, the TPA process occurs in the human eye, too, where two infrared photons may give rise to the sensation corresponding to that of light in the visible range \cite{palczewska2014,artal2017}. In this case the detection is not strictly PNR, as the signals from two infrared photons or from one visible photon are the same.

A process related to TPA is called stepwise two-photon absorption where the first photon takes the molecule to an actual (rather than a virtual) excited state and a subsequent photon takes the molecule to an even higher lying excited state, see e.g., \cite{kobayashi2018stepwise}.
Taking a molecule to an excited state, however,  is not yet sufficient for implementing a measurement. We also need an amplification process
that produces a macroscopic signal. In the human eye a light-absorbing molecule decays from the excited state irreversibly to a metastable state, in which the shape of the molecule has changed. That change in shape triggers a chain reaction of shape changes in surrounding proteins, eventually producing (or changing) a permanent dipole moment that in turn triggers a change in a mesoscopic electric current \cite{hall2020}, which then permanently registers the detection of the photon.

Following the example of Refs~\cite{young2018,young2018b,leonard2019,young2020design} of taking inspiration from biological systems to design photo detectors (see also \cite{chan2018}), based on this robust photo-detection mechanism  we propose and model a PNR two-photon detector 
consisting of a five-level molecule, as follows (see Fig.~1 and Section III for more details and reasons for choosing this particular configuration): a ground state $\ket{F_0}$ from which a photon with a frequency $\omega_\alpha\approx \omega_1-\omega_0=:\omega_{01}$ can induce a transition to an excited state $\ket{F_1}$, which can then irreversibly decay to a metastable state $\ket{F_2}$. In this state the molecule  triggers a first amplification process that indicates and permanently registers the detection of that first photon. Subsequently, a second photon of a different frequency $\omega_\beta\approx \omega_3-\omega_2=:\omega_{23}$ can excite the molecule to another state $\ket{F_3}$, from which it can decay to a different metastable state $\ket{F_4}$, triggering a second (different) amplification process that indicates the detection of the second photon.  The two photons must arrive sequentially rather than simultaneously for TPA.

One motivation for this work comes from recent theory efforts to find fundamental (i.e., device-independent) limits to photo detection \cite{vanenk2017,young2018,young2018b,yang2019,yang2019b,propp2019,proppb,biswas2020,propp2020}. For reasons fully explained in \cite{biswas2020} we construct a Hamiltonian here for the {\em full} detection process, including the crucial amplification step. A second motivation is of a more technical nature. The theoretical description of two (or more) photons interacting with a quantum system is known to be considerably more complicated than that of just a single photon interacting with the same system \cite{gheri1998,baragiola2012n,shi2015,nysteen2015,pan2016,baragiola2017,konyk2017,rag2017,dkabrowska2019b,molmer2019,molmer2020}.
Two types of methods have been developed to tackle this problem. One is based on a hierarchy of coupled differential equations for generalized density matrix elements \cite{gheri1998,baragiola2012n,mollow1975} for a quantum system interacting with prescribed multi-photon pulses. The other method \cite{molmer2019,molmer2020} includes virtual cavities that generate the photons and is thus based on a Hamiltonian description of the quantum system {\em and} the photons. We refer to these two methods as the ``generalized density matrix'' and the ``Hamiltonian'' formulations, respectively.
We will use both methods here, since they each have their own advantages, and we also give the explicit equations  (which seem not to have been given before) that link the two methods. 
Moreover, we can explain why the methods above yield expressions for scattered light and for the dynamics of the quantum system in terms of (Hilbert space) inner products that involve the temporal amplitudes of the incoming photons \cite{roulet2016} on the one hand, and the appropriate response functions of the system on the other.

This paper is organized as follows.  In Section II we first give a synopsis
of some of the results, which can be understood without going into the details of the derivations. Such details are provided in the remaining Sections. In Section III we give the Hamiltonian for our 5-level molecule.
Section IV describes the two different methods we used to obtain results:
the generalized density matrix methods is used to obtain  analytical results, while the Hamiltonian method is used to obtain numerical results. We explain why the latter method is so much easier to use for numerical calculations. Section V ends with conclusions and discusses possible extensions of our work.
In the Appendix we present the transformation that unifies the two formalisms (generalized density matrix and the Hamiltonian formulation) used in the paper. 

\section{Synopsis}\label{Synopsis}
Since the detailed description of our system is rather involved  we first give here a synopsis of the basic results without any derivations. The results presented here are quite straightforward to understand. 
\begin{figure}[h]\label{5level}
	\includegraphics[width=4in]{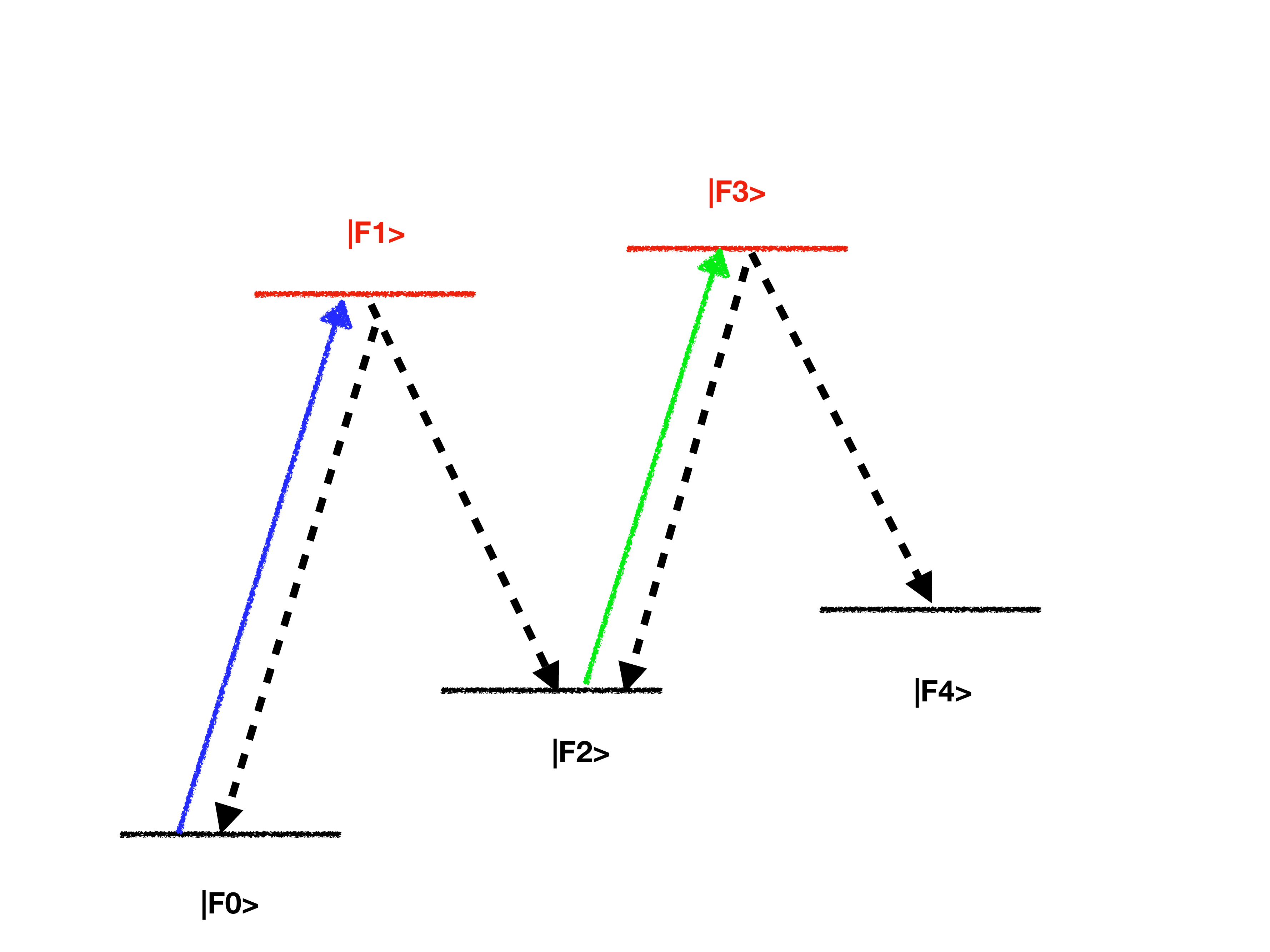}
	\caption{Model of a two-photon detector. A light-absorbing molecule starts in the ground state $\ket{F_0}$. There are two excited states $\ket{F_1}$ and $\ket{F_3}$, indicated in red, and two metastable states $\ket{F_2}$ and $\ket{F_4}$. We assume level $\ket{F_2}$ corresponds to a shape change [while still being an electronic ground state], which down the line corresponds to a change in electric dipole moment, which in turn induces a change in a mesoscopic current or voltage (thus mimicking the process taking place in the human eye). That mesoscopic change permanently registers the detection of the first photon.  The metastable state $\ket{F_4}$ corresponds to yet another change of shape, which eventually leads to a change in a dipole moment, which then in turn can change a mesoscopic current in a way that is distinct from what the molecule in state $\ket{F_2}$ accomplished. This distinct mesoscopic change then registers the second photon.
		The molecule can detect two photons, one ``blue'' photon resonant with the transition from the ground state to the first excited state, and a ``green'' photon  resonant with the transition from $\ket{F_2}$ to the second excited state. From each of the two excited states the molecule can spontaneously decay back to state it came from or to the desired metastable state. Thus there are four decay rates, indicated from left to right by $\gamma_1\ldots \gamma_4$, which are assumed to be more or less of the same order of magnitude. (The spontaneous transitions are indicated with dashed black lines.)
		On a time scale much longer than $1/\gamma_1$ the molecule resets by the metastable states decaying back to the ground state $\ket{F_0}$ (this resetting is not indicated in the figure).}
\end{figure}
The light-absorbing molecule at the heart of our detector is described in detail in Fig.~1. 
\subsection{Detection probabilities}

For an incoming single-photon wave packet, the different frequency components are not all absorbed with 100\% efficiency. The probabilty $P_\alpha$ for the first photon, labeled $\alpha$, to be detected can be written in the form
\be
P_\alpha=\int\! d\omega\,|T_1(\omega)|^2 |u_\alpha(\omega)|^2.
\ee
Here $u_\alpha(\omega)$ is the Fourier component of the incoming wave packet at frequency $\omega$ and may also be referred to as its spectral amplitude. $T_1(\omega)$ is a complex transmission amplitude for the molecule to go from the initial state $\ket{F_0}$ to the desired state $\ket{F_2}$ through the intermediate excited state $\ket{F_1}$ (see Eq.~(\ref{Tom}) below):
 \be\label{Tomsyn}
T_1(\omega)=\frac{\sqrt{\gamma_1\gamma_2}}{(\gamma_1+\gamma_2)/2-i\{\omega- \omega_{01}  \}}.
\ee
If $\gamma_1=\gamma_2$, the transmission probability $|T_1(\omega)|^2$ reaches a maximum of 1 at the resonance frequency $\omega_{01}$ and has a width of about $\gamma_1$. Thus, a resonant photon with a narrow width in frequency space (much less than $\gamma_1$) and whose duration is, therefore, much longer than $\gamma_1^{-1}$, can be absorbed with near-unit efficiency, exactly as was found before in Refs.~\cite{young2018,young2018b,propp2019}.

A similar result holds for the second photon, labeled  $\beta$. The only (important!) difference is that the second photon can be absorbed only when the molecule is in the state $\ket{F_2}$. Hence ideally it should arrive after photon $\alpha$ has been fully absorbed. In that ideal case, the conditional probability of detecting photon $\beta$ (with a spectral amplitude $u_\beta(\omega)$), given that photon $\alpha$ was detected, is
\be\label{Pb}
P_\beta=\int d\omega |T_2(\omega)|^2 |u_\beta(\omega)|^2,
\ee
with
 \be\label{Tom2syn}
T_2(\omega)=\frac{\sqrt{\gamma_3\gamma_4}}{(\gamma_3+\gamma_4)/2-i\{\omega- \omega_{23}  \}}
\ee
a second complex transmission amplitude, describing how the molecule can transition from level $\ket{F_2}$ to level $\ket{F_4}$ through the intermediate $\ket{F_3}$ excited state.

The probability to detect both photons in the more general case when the two photons do overlap in time can be written in the form
\be
P_{\alpha\&\beta}=P_\alpha P_\beta -P_{{\rm overlap}},
\ee
where the (non-negative) ``overlap term'' will be derived and discussed in Section \ref{GDMO}. We merely note here that the overlap term can be found analytically and is then written as a convolution involving the two spectral amplitudes $u_{\alpha,\beta}(\omega)$ and the two transmission amplitudes $T_{1,2}(\omega)$. If photon $\beta$ is delayed by a time much longer than $1/\gamma_1$, then $P_\overlap\rightarrow 0$, but if photon $\beta$ entirely precedes photon $\alpha$, then $P_\overlap\rightarrow P_\alpha  P_\beta$. 

\begin{figure}[h]
	\includegraphics[width=4.9in]{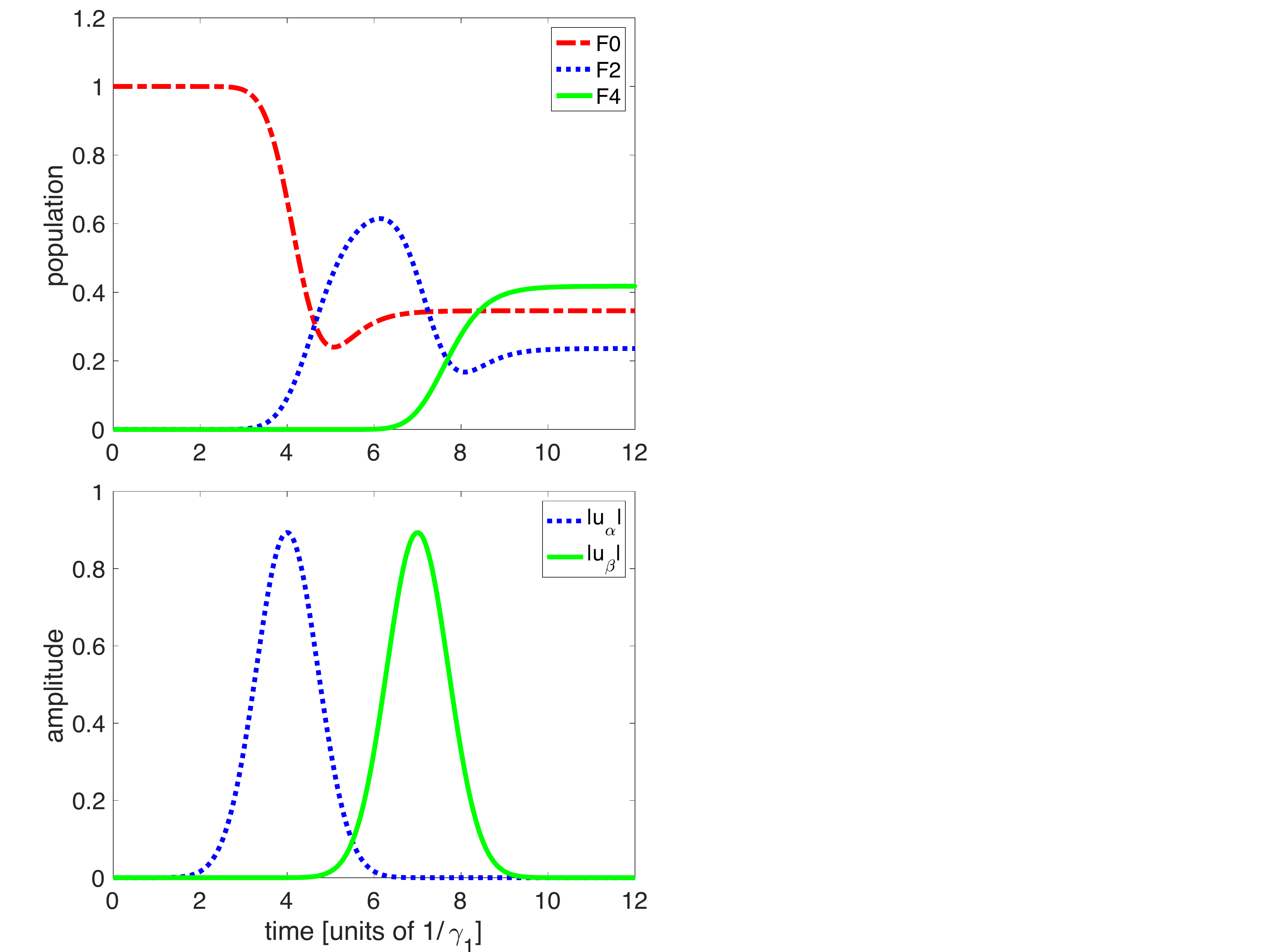}
	\caption{Top: Populations in the ground state and the two metastable states as functions of time, when two photons arrive sequentially. Bottom: the (Gaussian) amplitudes of the ``blue'' photon ($u_\alpha$) and the ``green'' photon ($u_\beta$) as functions of time. We chose here $\gamma_k=\gamma_1$ for $k=2,3,4$ and the time delay between the two input photons is $3/\gamma_1$
		Eventually a steady state is reached, with the total population in the three lowest states adding up to 1. The steady-state population in the ground state (dot-dashed curve) is 0.346, which equals the probability to not detect any of the photons. The sum of the steady-state populations in the metastable states is .654 and equals the probability to detect the ``blue'' photon.
		The steady-state population in $\ket{F_4}$ is 0.418 and equals the probability to detect both photons.}
\end{figure}
In Figure 2 we plot a numerical result for a case that is not optimal for two reasons. First, the widths in time of the two incoming single-photon pulses are equal to $1/(2\gamma_1)$, which is too short to be close to optimal. Second, the pulses partially overlap in time. The probability to detect both photons is then about 42\%.

\begin{figure}[h]
	\includegraphics[width=4.9in]{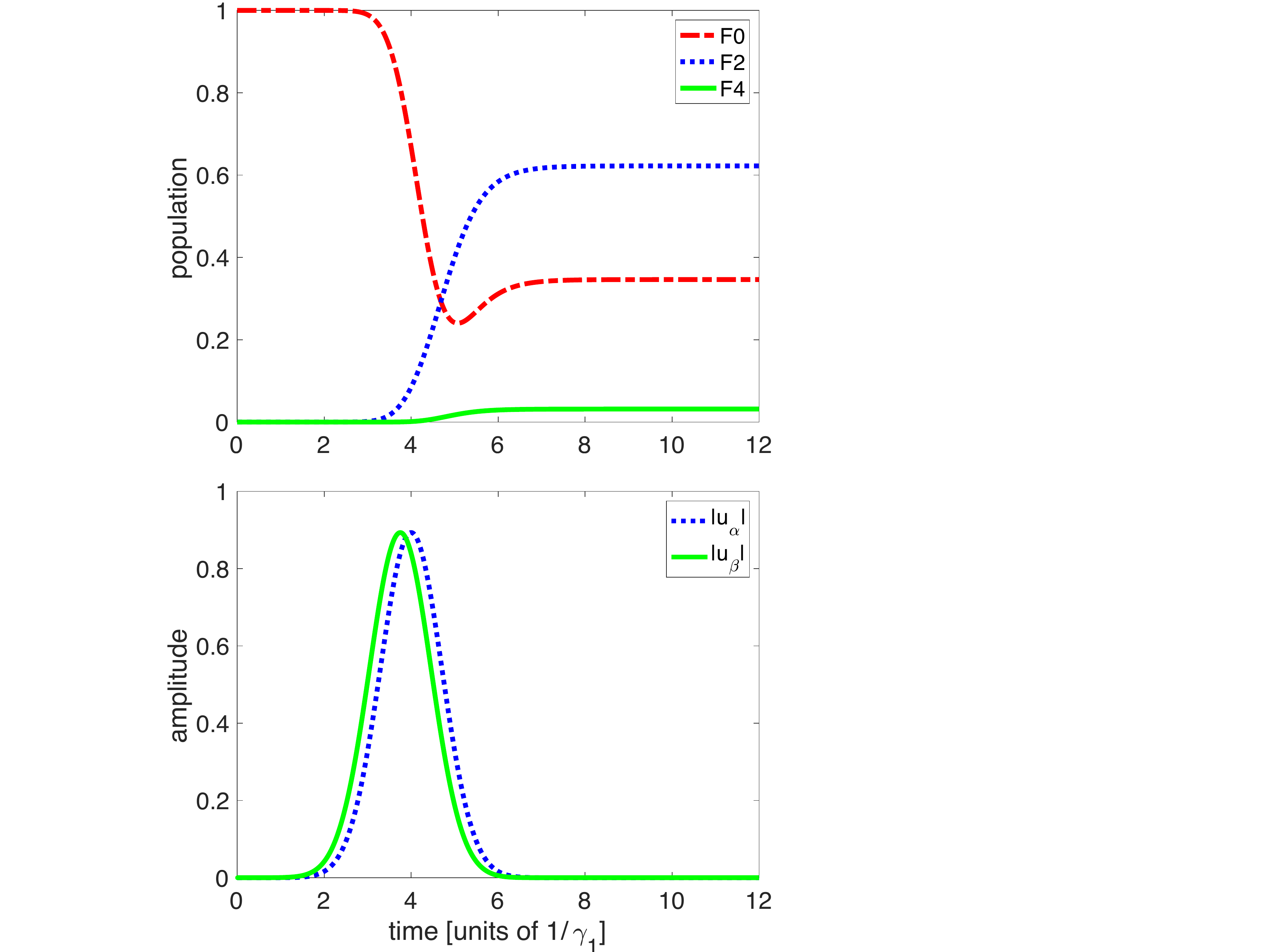}
	\caption{Top: Populations in the ground state and the two metastable states as functions of time, when the ``green'' photon arrives just before the ``blue'' photon (the time delay is $-1/(4\gamma_1$). Bottom: the absolute values of the amplitudes $|u_\alpha|$ of the ``blue'' photon  and $|u_\beta|$ of the ``green'' photon  as functions of time.
	The probability to detect both photons is very small in this case, 0.022. The first photon is detected with the same probability (0.346) as in the previous figure.}
\end{figure}

In Figure 3 we plot a case that shows how important the delay between the two photons is. Here photon $\beta$ arrives just before photon $\alpha$: while this does not affect at all the absorption (and detection) of photon $\alpha$, photon $\beta$ is now detected only with a very small probability of about 2\%.
\subsection{Detector clicks}
The generalized density matrix formalism can be used to get analytical expressions  describing ``clicks'' of our detector in simple cases.
\subsubsection{One photon}
For example, suppose for simplicity that we could measure in what state our molecule is at a specific time $T>t_0$, given that it started in the state $\ket{F_0}$ at time $t_0$, and suppose that we find our molecule in the state $\ket{F_2}$. This clearly would implement a measurement of the incoming photon. Thus, ignoring the second photon for now, given an expression for the population in that level as a function of time, we can write that probability at time $T$ in the form of the Born rule as
\be \label{uPi}
P_2(T)=\Tr (\proj{u_\alpha} \Pi_1),
\ee
where $\proj{u_\alpha}$ is the projector onto the input single-photon wave packet of photon $\alpha$, and $\Pi_1$ is a positive hermitian operator (guaranteeing that $P_2(T)$ is a real non-negative number). 
We can always write $\Pi_1$ in a diagonal form
\be\label{diago}
\Pi_1=\sum_n \lambda_n \proj{\phi_n}
\ee
with $\lambda_n$ real and non-negative, and with $\{\ket{\phi_n}\}$ forming an orthonormal basis of single-photon states. That means the probability $P_2(T)$ can be rewritten as
\be
P_2(T)=\sum_n \lambda_n |\braket{u_\alpha}{\phi_n}|^2.
\ee
The fact that the Born rule is linear in the input state (represented as a density operator or matrix) thus explains why this probability can be expressed in terms of overlaps involving the incoming single-photon wave packet \cite{roulet2016,muller2017}. It also follows that $\lambda_n\leq 1$, since $\lambda_n$ has the meaning of the probability that an input photon in the state $\ket{\phi_n}$ will be detected.

In our specific case we find that $\Pi_1$ is of the form
\be\label{POVMWt}
\Pi_1=\int_{t_0}^T\! dt \,
W_t \proj{\phi_t}
\ee
where $W_t$ is a weight per unit of time
\bea\label{Wt}
W_t=\frac{\gamma_1\gamma_2}{\gamma_1+\gamma_2}
\left[1-\exp(-(\gamma_1+\gamma_2)(t-t_0))\right],
\eea
and the projector projects onto a normalized single-photon state of the form
\be\label{phit}
\ket{\phi_t}= \frac{\int_{t_0}^t dt' \exp[(\gamma_1+\gamma_2)t'/2]\exp(i\omega_{01} (t'-t)) b_1^\dagger (t') \vac}{\sqrt{\int _{t_0}^t dt' \exp[(\gamma_1+\gamma_2)t']}}
\ee
where $b_1^\dagger(t)$ is the Fourier transform of $b_1^\dagger(\omega)$. (Note that we could equivalently write $t'-t$ instead of $t'$ in the arguments of the $\gamma_{1,2}$-dependent exponentials in both numerator and denominator.) These states $\ket{\phi_t}$ are not orthogonal for different values of $t$ and this type of nonorthogonal states also appears in the context of spectral filtering \cite{vanenk2017b}. We also note that the transmission function $T_1(\omega)$ given above in Eq.~(\ref{Tomsyn}) is the (properly normalized) Fourier transform of the time-dependent function ---which is a Green's function---appearing in $\ket{\phi_t}$. That transmission function also determines the spectral shape of the photon emitted spontaneously by the molecule \cite{muller2017}.

It is important to note that in Eq.~(\ref{uPi}) $\Pi_1$ refers only to the detector, and $\ket{u_\alpha}$ refers only to the incoming photon. $\Pi_1$ is called a POVM (Positive-Operator Valued Measure) element and fully describes the outcome of the measurement corresponding to finding the molecule in level $\ket{F_2}$ at time $T$. It allows us to calculate for {\em any} incoming photon the detection probability (\ref{uPi}). In particular, it allows us in principle to infer the type of photon that is detected with the largest possible probability, by making use of the diagonal form (\ref{diago}). The largest eigenvalue $\lambda_{\max}=\max_n \lambda_n$ gives the highest possible efficiency $\eta_{\max}=\lambda_{\max}$ of detecting a single photon,  and the corresponding eigenstates [there may be more than one] give the optimal single-photon states that achieve that limit.

The interpretation of
\bea
\Tr(\Pi_1)&=&\sum_n \lambda_n=\int_{t_0}^{T}\! dt\, W_t
\nonumber\\
&\approx&  \frac{\gamma_1\gamma_2}{\gamma_1+\gamma_2}
\left[T-t_0-\frac{1}{\gamma_1+\gamma_2}\right],
\eea
(where we ignored an exponentially small term in the second line) is that of a bandwidth: the effective size of the single-photon Hilbert space covered by this particular measurement outcome \cite{vanenk2017}. This bandwidth may be (much) larger than unity. For a fixed value of $\gamma_1+\gamma_2$ the bandwidth is maximized by $\gamma_1=\gamma_2$, an optimal  ``impedance-matching'' condition found before in the same context of designing an optimal single-photon detector \cite{young2018,young2018b,propp2019}. The bandwidth is then approximately equal to the total time the detector has been on in units of $2/\gamma_1$. 

If we would be able to measure if the molecule were in state $\ket{F_1}$ at time $t$, then the corresponding POVM element would be proportional to a pure projector. But, since we do not know when the upper state spontaneously decayed to state $\ket{F_2}$, we do not know $t$, and hence we get a mixed POVM element. That is, for fixed $T$ (when we detect the molecule to be in the state $\ket{F_2}$) there are different possibilities for time $t$, each with their own probability $W_t dt$. That is  the interpretation of (\ref{POVMWt}). 

The idea that a quantum system absorbs a single-photon wave packet with in principle 100\% efficiency if and only if it is the time-reversed version of a photon that the system would emit if it started in the final state \cite{stobinska2009,raymer2018} does not apply so simply here, because of the presence of irreversible spontaneous decay. If we imagine we would apply a laser pulse to the $\ket{F_1}\rightarrow\ket{F_2}$ transition to induce stimulated emission, then, as is well known \cite{gorshkov2007universal,giannelli2018}, that idea indeed would apply straightforwardly . 
 \subsubsection{Two photons}
	The more interesting case of detecting the molecule in level $\ket{F_4}$ at time $T$ signals the detection of both photons and is described by the POVM element
\be\label{POVM2}
\Pi_2=   \int_{t_0}^T\! dt \, \int_{t_0}^t \! dt'\,
W_{t'} W_{t,t'}  \proj{\phi_{t'}}\otimes \proj{\psi_{t,t'}},
\ee
with
\bea
W_{t,t'}=\frac{\gamma_3\gamma_4}{\gamma_3+\gamma_4}
\left[1-\exp(-(\gamma_3+\gamma_4)(t-t'))\right],
\eea
and the single-photon state corresponding to the second photon is
\be
\ket{\psi_{t,t'}}= \frac{\int_{t'}^{t} d\tau \exp[(\gamma_3+\gamma_4)\tau/2]\exp(i\omega_{23} (\tau-t)) b_2^\dagger (\tau) \vac}{\sqrt{\int _{t}^{t'} d\tau \exp[(\gamma_3+\gamma_4)\tau]}}.
\ee
The prefactor $W_t$ and the single-photon state $\ket{\phi_t}$ appearing here are exactly as defined before in (\ref{Wt}) and (\ref{phit}).
The time-dependent function appearing in $\ket{\psi_{t,t'}}$ is once again a Green's function, and $T_2(\omega_b)$ is its (normalized) Fourier transform.

 There is a double integral over time in (\ref{POVM2}), each integral corresponding to an irreversible step in the detection process, which makes it uncertain at what time $t$ we could have found the molecule in state $\ket{F_3}$ and at what earlier time $t'<t$ we could have found the molecule in $\ket{F_1}$.

We may again write down an eigenvalue equation for $\Pi_2$ [which would have to be solved numerically] and then write that POVM element in the diagonal form
\be
\Pi_2=\sum_n \mu_n \proj{\phi_n^{(\alpha,\beta)}},
\ee 
where the projectors $\proj{\phi_n^{(\alpha,\beta)}}$ project onto specific pure two-photon (eigen)states, and the eigenvalues $0\leq \mu_n\leq 1$ give the corresponding efficiencies with which those specific two-photon wave packets are detected at time $T$.

Like we saw for the single-photon case treated above, the bandwidth 
\be
\Tr(\Pi_2)=\sum_n \mu_n= \int_{t_0}^T\! dt \, \int_{t_0}^t \! dt'\,
W_{t'} W_{t,t'} 
\ee
 is the size (dimension) of the two-photon Hilbert space covered by our detector.
\section{The two photon absorber and its Hamiltonian}\label{Ham2}
To construct the minimal absorber atom or molecule or multi-level system that can absorb two photons sequentially and produce classical outputs signaling the final state of the absorber, we consider the five level system of Figure 1 for efficient photon transduction. Some recent efforts for physically based fundamental models for photo detection assemble all parts of the process into a single fully coupled evolution problem  \cite{vanenk2017,young2018,young2018b,yang2019,yang2019b,propp2019,proppb,propp2020,biswas2020,propp2020}. Minimal noise amplification of the absorbed photon signal has been shown to be optimally done with continuous quantum measurement \cite{young2018,young2018b,biswas2020}. In this scheme, the ``shelving state'' or the state in which the absorber produces the amplified classical readout is continuously measured. To get around the quantum Zeno effect problem with having the same state to be the photo-excited and shelving state, a three level system is determined to be optimal for single photon detection \cite{young2018}. Hence we use the three levels $\ket{F_{0,1,2}}$ to detect one photon.

For the two-photon detection scheme, we supplement the molecule with two more levels. The second photon can lift the molecule from state $|F_2\rangle$ into the excited state $|F_3\rangle$ which can spontaneously relax into the second shelving state $|F_4 \rangle$. In the latter state the molecule  triggers an amplification process which produces  a noticeably different signal than that produced by the shelving state $|F_2\rangle$. The absence of a signal and the two different signals from the two levels $|F_2\rangle$ and $|F_4\rangle$ help the observer distinguish the number of photons (0, 1, or 2) absorbed by the molecule. Since the frequency of the amplified signal is independent of the input photon frequency \cite{proppb}, we can have different shelving states (classically) driving different oscillators of different frequencies \cite{biswas2020}; and hence we can have distinguishable classical output signals for one or two detected photons.

We wish to calculate the dynamics of the 5-level discrete quantum system F coupled to the two continua $b_1$ and $b_2$ which contain our two input photons (with different frequencies). With $\hbar=1$, the parts of the Hamiltonian in the Markov approximation for these coupled systems are 
\bea
H_{sys} = \sum\limits_{k=0}^{4} \omega_k |F_k \rangle \langle F_k|,  \ \ \ \ \ \ \ \ \ \ \label{eq_HF}
\eea
\bea
H^{1(2)}_{bath} = \int d\omega \ \omega b_{1(2)}^{\dagger}(\omega) b_{1(2)}(\omega),  \ \ \ \ \ \label{eq_Hbath}
\eea

\bea
H_{int} = - i \int d\omega \left[ \sqrt{\frac{\gamma_1}{2\pi}} |F_1 \rangle \langle F_0|  b_1(\omega) \right. \nonumber\\
\left. +\sqrt{\frac{\gamma_3}{2\pi}} |F_3 \rangle \langle F_2| b_2(\omega)  \right]+ H.c.  \ \ \ . \ \label{eq_Hint}
\eea
This part of the Hamiltonian includes spontaneous decay back to  $\ket{F_0}$ and back to $\ket{F_2}$.
(The radiation field modes are fully described by four degrees of freedom. Here we fixed the quantum numbers for three of them (polarization and two transverse spatial degrees of freedom) and explicitly retain only the spectral/temporal degree of freedom.)

The next and last part of our Hamiltonian is necessary for the purpose of enabling the additional spontaneous decays of the absorber from $|F_1 \rangle$ to $| F_2 \rangle$ and from $|F_3 \rangle$ to $| F_4 \rangle$. These two transitions need to be dipole allowed and $\gamma_2$ and $\gamma_4$ determine the rates (probability per unit time) of those two processes:
\bea
H^1_{int} = - i \int d\omega \left[\sqrt{\frac{\gamma_2}{2\pi}} |F_1 \rangle \langle F_2|  g(\omega) \right. \nonumber\\
\left. +\sqrt{\frac{\gamma_4}{2\pi}} |F_3 \rangle \langle F_4| h(\omega)  \right] + H.c. \ \ \ . \ \label{eq_Hint_decay}
\eea
in terms of two additional independent (commuting) bosonic modes, described by annihilation operators $g(\omega)$ and $h(\omega)$ and their hermitian conjugates.
\section{Two theories for photon absorption}\label{Theories}
Restriction of the number of excitations to one or two offers a workaround for the complications of the multimode nature of the interaction of  propagating light with a nonlinear medium such as a two- or three-level atom. The Fock state master equation formalism by Baragiola \textit{et. al.} \cite{baragiola2012n}, and the set of generalized density matrices by Gheri \textit{et. al.} \cite{gheri1998} offer suitable theoretical frameworks for calculating few photon Fock state interactions with a multi-level discrete quantum system. 

An alternate route for having a computationally manageable effective master equation has recently been developed by Kiilerich \textit{et. al.} \cite{molmer2019,molmer2020} by restriction of the input pulse to a single time dependent mode. This approach is appealing to the problem of single photon absorption as the same physical effects of the incoming wave packet of the multimode bosonic input field is emulated. As previously formulated by Gheri \textit{et. al.} \cite{gheri1998}, an upstream virtual cavity is introduced whose output serves as the incident field for a system under study. The incident field generated by the cavity is in a state residing in a specific wave-packet mode and all other orthogonal modes are designated the vacuum state. Since we are only interested in the input quantum state and the absorption of the photon, we only acquire the technique of {\em driving} with a quantum pulse from Ref. \cite{molmer2019,molmer2020}. The {\em reflected} quantum state is of no interest to us, and only the transmitted state (\cite{propp2019}) which quantifies the probability of absorption is required for our purpose. 

The generalized density matrices framework developed by Gheri \textit{et. al.} \cite{gheri1998} suffices for calculating the absorption probabilities and corresponding POVMs. However, we introduce the virtual upstream cavities and formulate a Hamiltonian formulation for the entire evolution problem of photo detection (including amplification to a mesoscopic signal) that we introduced in the previous publication \cite{biswas2020}. The Hamiltonian formulation is versatile and facilitates the calculations to be done in either the Schr\"odinger or the Heisenberg picture. The explicit transformation between the generalized density matrices and the components of the density matrix obtained by the Hamiltonian method is presented in Appendix A.

\subsection{Generalized Density Matrix Operators}\label{GDMO}
We assume we have two unentangled single-photon wavepackets in two orthogonal modes
\bea \label{eq_psi2}
| \Psi_{2} \rangle = | \Psi_{\alpha} \rangle | \Psi_{\beta} \rangle, 
\eea
where the individual photon states are defined as
\bea
|\Psi_{\alpha(\beta)} \rangle = \int\limits_{-\infty}^{\infty}d\omega_{a(b)}  u_{\alpha(\beta)}(\omega) b_{1(2)}^{\dagger}(\omega) \vac.
\eea
$u_{\alpha(\beta)}$ is the properly normalized wave function for  photon $a$ ($b$). The two photons reside in the two distinct continua $b_1$ and $b_2$. (We will also use the Fourier transforms of the single-photon amplitudes, which for simplicity we denote by $u_{\alpha,\beta}(t)$.)

Following Ref. \cite{gheri1998}, we can define generalized density matrix operators for $i,j=0,2,\alpha,\beta$ and derive a set of coupled differential equations for them that describes the absorption of the two photons. In the following, $R$ denotes the reservoir or bath, which includes continua other than $b_1$ and $b_2$, such as the continua $g$ and $h$ introduced above:
\bea
\mathcal{\rho}_{i,j}(t)= \Tr_R \left[U(t,t_0)\mathcal{\rho}_S(t_0) \otimes | \Psi_i \rangle \langle \Psi_j | U^{\dagger}(t,t_0) \right]. \ \ \ \ \label{eq_gen_den}
\eea
Here $| \Psi_0 \rangle $ denotes the vacuum state $\vac$, and $| \Psi_2 \rangle $, $| \Psi_{\alpha} \rangle $ and $| \Psi_{\beta} \rangle $ are the two-photon input state and the individual single-photon states introduced above. Furthermore, $\rho_S(t_0)$ is the initial state of all remaining quantum systems, including our 5-level molecule and the reservoir $R$.  In our case, each of these generalized density matrices for fixed values of $i$ and $j$ is a 5x5 matrix, describing the 5 levels of our molecule.

The generalized density matrices can be expanded in a time independent complete 5x5 basis, and substitution in the evolution equations yields a set of coupled differential equations for the coefficients $\mathcal{\rho}_{ik,jl} (t)$ of the expansion,
\bea
\mathcal{\rho}_{i,j}(t) = \sum\limits_{k,l} \mathcal{\rho}_{ik,jl} (t) | F_k \rangle \langle F_l|. \ \ \ \label{eq_exp_F}
\eea
These equations are given in Appendix A.
The diagonal generalized density matrices (for $i=j$) have a preserved trace of 1, and off-diagonal ones have a preserved trace of 0 over the evolution \cite{gheri1998}.  (In the alternative Hamiltonian formulation shown below a single Hamiltonian (with auxiliary cavities appended) can embody the complete evolution, and a {\em single} density matrix  (with preserved trace of 1) of size 20x20 can embody the complete dynamics \cite{biswas2020}.)

In order to simplify intermediate equations, we will absorb a time-dependent phase factor $\exp(i\omega_{01})t)$ in the definition of the single-photon amplitude $u_\alpha(t)$ for photon $a$ and similarly a factor $\exp(i\omega_{23}t)$ in the amplitude $u_\beta(t)$ for photon $b$, such that both amplitudes can be considered slowly-varying if the photons are more or less on resonance with their respective transition in the molecule. End results are quoted in terms of the original amplitudes.

The evolution problem is initiated with $\mathcal{\rho}_{00,00}=1$ at time $t_0$, i.e., the molecule is in the $|F_0 \rangle$ state, with any photon yet to come in. The coefficient $\mathcal{\rho}_{\alpha2,\alpha2}$ embodies the evolution of the molecule occupation elevated to $|F_2 \rangle$ state driven by just the first photon $\alpha$ with temporal amplitude $u_\alpha(t)$. The solution found is
\bea
 \mathcal{\rho}_{\alpha 2,\alpha 2}(t) = \gamma_1 \gamma_2 \int\limits_{t_0}^t dt_1 \left[ e^{-(\gamma_1+\gamma_2) t_1}  \right. \ \ \ \ \ \ \ \ \ \ \ \ \ \ \ \ \ \ \ \ \ \ \nonumber\\
\left. \int\limits_{t_0}^{t_1} dt_2 e^{\frac{\gamma_1+\gamma_2}{2}t_2}  u^*_{\alpha}(t_2) \int\limits_{t_0}^{t_2} dt_3 e^{\frac{\gamma_1+\gamma_2}{2}t_3}  u_{\alpha}(t_3)  \right] + c.c., \ \ \ \ \ \ \label{eq_a2a2}
\eea
This result becomes especially simple when considering the steady-state, obtained by taking the limit $t \rightarrow \infty$. The result further simplifies when we take the limit $t_0\rightarrow-\infty$ such that in principle any single-photon wave packet could be absorbed, irrespective of when it arrives. The equations in those limits are most easily solved in Fourier space,  and we obtain then the same result we had obtained before in Ref.~\cite{biswas2020},
 \be\label{uTo}
  \mathcal{\rho}_{\alpha 2,\alpha 2}(\infty) = P_{\alpha}=\int\!d\omega\,
  |u_{\alpha}(\omega)|^2|T_1(\omega)|^2,
 \ee
 where  
 \be\label{Tom}
 T_1(\omega)=\frac{\sqrt{\gamma_1\gamma_2}}{(\gamma_1+\gamma_2)/2-i\{\omega- (\omega_{1}-\omega_{0})  \}}
 \ee
 is the transmission coefficient describing the propagation of a single excitation through the  $\Lambda$ system \cite{propp2019,vanenk2017b}. (This is Eq.~(\ref{Tomsyn})) of the Synopsis Section.)

$\mathcal{\rho}_{\alpha 2,\alpha 2}(t)$ in Eq. \ref{eq_a2a2} can be recast into the more informative form
	\bea \label{r22}
	\mathcal{\rho}_{\alpha 2,\alpha 2}(t) = \int\limits_{t_0}^t dt' \left| \int_{t_0}^{t'}\!dt_2\, \sqrt{\gamma_1 \gamma_2 }  e^{\frac{\gamma_1+\gamma_2}{2}(t_2-t')}   
	u_{\alpha}(t_2)   \right|^2. \nonumber\\
	\ \ \ \ \ \ \label{eq_a2a2_1}
	\eea
This is the form that can be used straightforwardly to obtain the expressions (\ref{POVMWt})--(\ref{phit}) for the POVM element $\Pi_1$.
The quantity inside the integral over $t'$ is actually $\gamma_2$ times the population in level $\ket{F_1}$ as a function of time \cite{young2018b}.

The probability of the molecule reaching $|F_4 \rangle$ state driven by the second photon $\beta$ has a ``nested'' structure containing the expression, $\mathcal{\rho}_{\alpha 2,\alpha 2}(t)$,
\bea
 \mathcal{\rho}_{24,24}(t) = \gamma_3 \gamma_4 \int\limits_{t_0}^t dt_1 \left[ e^{-(\gamma_3+\gamma_4) t_1}  \right. \ \ \ \ \ \ \ \ \ \ \ \ \ \ \ \ \ \ \ \ \ \ \nonumber\\
\left. \int\limits_{t_0}^{t_1} dt_2 e^{\frac{\gamma_3+\gamma_4}{2}t_2}  u^*_{\beta}(t_2) \int\limits_{t_0}^{t_2} dt_3 e^{\frac{\gamma_3+\gamma_4}{2}t_3}  u_{\beta}(t_3) \mathcal{\rho}_{\alpha 2,\alpha 2}(t_3) \right] + c.c. \nonumber\\
\label{eq_2424}
\eea
We may rewrite this expression by changing variables in the complex conjugate term and by substituting (\ref{r22}) to obtain our two-photon POVM (\ref{POVM2}).

As we noted in Section \ref{Synopsis} the time-dependent functions appearing in the expression for $\rho_{24,24}(t)$ and other populations of quantum levels can be interpreted as Green's functions. Their (normalized) Fourier transforms act as transmission and reflection coefficients when treating this problem as a scattering problem. In our case, transmission coefficients $T_1(\omega)$ (defined above) and $T_2(\omega)$ (defined below) play the new role of determining the detection probability of photons with frequency $\omega$ in the limit of $t\rightarrow\infty$, as we saw in Eq. (\ref{uTo}) and as we will show in the next subsection.

\subsection{Overlap term}\label{Overlap}
The absorption of the two photons can be completely calculated in the frequency domain. To that end, we define the Fourier transform of the population in level $\ket{F_1}$, since we can express all quantities of interest in terms of that function. We find 
\bea
F_1(\omega) =\frac{1}{\sqrt{2\pi}}\int\! dx \,
T_1(\omega+x) u_\alpha (\omega+x)
T_1^*(x) u^*_\alpha (x). \ \ \ \ \ \ \ \
\eea
The 0-frequency component of $\sqrt{2\pi}F_1(\omega)$, equals the detection probability for the first photon $\sqrt{2\pi}F_1(0)=\int dx |u_\alpha |^2 |T_1(x)|^2=:P_\alpha$.
 In the frequency domain, we then obtain the Fourier transform of $\rho_{\alpha 2,\alpha 2}(t)$ as
\be
\mathcal{\rho}_{\alpha2, \alpha2}(\omega) =\gamma_2\left[ iF_1(\omega){\rm P} \left(\frac{1}{\omega}\right) + \pi F_1(0) \delta(\omega)\right].
\ee
where P denotes the principal value.
In eq.~(\ref{eq_2424}), if we replace $\mathcal{\rho}_{\alpha2,\alpha2}(t)$ with its steady state value $P_{\alpha}$, we get an expression identical in form to eq.~(\ref{eq_a2a2}) with different decay rates, and we thus can simply evaluate the result for $t\rightarrow\infty$  as the product $P_{\alpha}P_{\beta}$ with $P_\beta$ given by (\ref{Pb}). So, if the second photon arrives long after the first photon has been completely absorbed (and the absorber raised to the level $|F_2 \rangle$), the probability of both photons being absorbed becomes the product of their individual absorption probabilities. 

Therefore, we can rewrite the probability of two-photon absorption, $\mathcal{\rho}_{24,24}(\infty)$ as a sum of two parts, one being the product of the two absorption probabilities. We name the other term $P_{\overlap}$, since we expect the term to vanish if the second photon comes in after a delay and the two wave functions of the two photons overlap negligibly. We thus write
\bea
\mathcal{\rho}_{24,24}(\infty) = P_{\alpha}  P_{\beta} - P_{\overlap} \ \ \ \ \ \ \ \ \ \ \label{eq_r2424}. 
\eea
After some algebra, we obtain
\bea
P_{\overlap} = \frac{1}{2} P_{\alpha}  P_{\beta} + P^{\alpha\beta} \eea
where
\bea
P^{\alpha\beta} &=& \frac{\gamma_2\sqrt{\gamma_3\gamma_4}}{\sqrt{2\pi}(\gamma_3+\gamma_4)}
\int\! d\omega\, 
 u^*_{\beta}(\omega) T_2(\omega)\nonumber\\
 &\times& 
 \int \! dx \, u_{\beta}(\omega-x)  {\rm P} \frac{F_1(x)}{ix} 
+ c.c.,  \nonumber\\
\label{eq_Pab}
\eea
where P denotes the principal value. The following results are borne out in numerical simulations for different arbitrary wave shapes of the two photons that are delayed by a long time $t_d>>1/\gamma_1$:
\bea
P^{\alpha\beta} &\longrightarrow& -\frac{1}{2} P^{\alpha}_{abs}  P^{\beta}_{abs} \nonumber\\
P_{\overlap}&\longrightarrow& 0 \nonumber\\
\mathcal{\rho}_{24,24}(\infty)&\longrightarrow&  P_{\alpha} P_{\beta}.
\eea

\subsection{Hamiltonian Formulation}

In a recent paper, Ref. \cite{biswas2020}, we developed a ``Hamiltonian formulation'' that can describe a single photon detection process in its entirety. We now adapt that formulation for the detection of two unentangled photons absorbed sequentially.  The most convenient method for solving the dynamical equation set is numerical integration of the Liouvillian equation in the Hamiltonian formulation \cite{biswas2020}. In the Hamiltonian formulation, we get a single density matrix for the entire system that can be solved easily with well-known vectorization and Trotter decomposition techniques \cite{granade2015characterization}. From the solution of the single density matrix, the generalized density matrices can be found easily with the transformation (\ref{eq_trans}).

 We introduce two auxiliary cavities with damped harmonic motion leaking one excitation each into the continuous  bath modes $b_1(\omega)$ and $b_2(\omega)$. These two excitations mimic the photon wave packets in the two baths that we are trying to detect. There are two other continuous modes $g$ and $h$, which are introduced only to enable the spontaneous relaxation of the molecule F.
 \begin{center}
	\begin{figure}
		\includegraphics[width=3.4in]{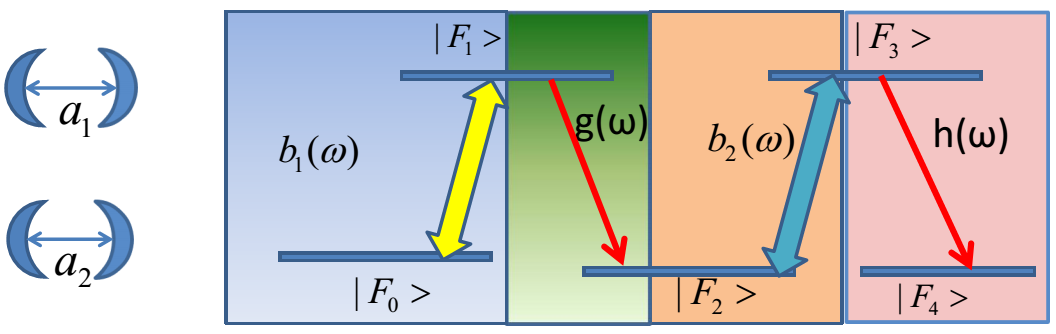}
		\caption{The cavity modes $a_1$ and $a_2$ each have one excitation to start with. These two excitations leak out into their adjacent baths (continuous modes $b_1$ and $b_2$ respectively) by designing the coupling to the baths in time, thus creating two single-photon wavecpackets. They respectively drive the $|F_0\rangle$ to $|F_1 \rangle$ and $|F_2\rangle$ to $|F_3 \rangle$ transitions. From the excited levels, the molecule can relax with certain probabilities either back to the state it came from or to another shelving state. The two shelving states drive two distinct amplification processes and thus produce two macroscopically distinct ``classical'' signals (unrelated in frequency to the incoming photons) in an output bath $d(\omega)$.}
	\end{figure}
\end{center}
The Hamiltonian is of the following form
\bea\label{H}
H=H_{a_1}+H_{a_1-b_1}+ H_{b_1}+H_{a_2}+H_{a_2-b_2}+ H_{b_2}\nonumber\\
+H_{b_1-F}+H_{b_2-F}+\nonumber\\
H_F+H_{F-g}+H_g+H_{F-h}+H_h\nonumber\\
+H_{F-c}+H_c+H_{c-d}+H_d.
\eea
The diagonal terms in the Hamiltonian give all the eigen energies of the  systems. For example, for the cavities it features their resonance frequencies, 
\bea
H_{a_{1(2)}} &=& \omega_{a_{1(2)} }a_{1(2)}^{\dagger}a_{1(2)},
\eea
and for the continuous modes, such as $g$,
we have
 \bea
H_g &=& \ido \omega g^{\dagger}(\omega)g(\omega),
\eea
and similar terms for $H_{b_1}, H_{b_2}, H_c, H_d$ and $H_h$. $H_F$ is simply $H_{sys}$ as defined before in Eq.~(\ref{eq_HF}). 

The interaction between the cavities and the field modes, as well as the interaction of the photons with the molecule are mediated by the electric fields corresponding to modes $b_1$ and $b_2$.
Each of the electric field operators of the modes can be expanded into plane wave basis (also their Hermitian conjugate operators). For the input fields $B_{1,2}(x,t)$, we expand
\bea\label{Bx}
B_{1,2}(x,t)&=&\frac{1}{\sqrt{2\pi}}\ido b_{1,2}(\omega,t)\exp(i\omega x/c).
\eea
The molecule is located at $x=0$ and
the cavities $a_1$ and $a_2$ are located ``upstream'' at $x= -c \tau_1$ and $x= -c \tau_2$ where $c$ is the speed of light and $\tau_1, \tau_2$ are the times it takes for a photon to travel from the respective cavities to the absorber F. The cavities are coupled to the fields $B_1(x=-c\tau_1,t)$ and $B_2(x=-c\tau_2,t)$in the manner:
\bea
H_{a_{1(2)}-b_{1(2)}}= \ \ \ \ \ \ \ \ \ \ \ \ \ \ \ \ \ \ \ \ \ \ \ \ \ \ \ \ \ \ \ \ \ \ \ \ \ \ \ \ \ \ \ \ \ \ \ \ \ \ \ \ \ \ \ \nonumber\\
i  [g^*_{1(2)}(t)a_{1(2)}B_{1(2)}^{\dagger}(-c\tau_{1(2)},t)-g_{1(2)}(t)B_{1(2)}(-c\tau_{1(2)},t)a_{1(2)}^{\dagger}] \nonumber
\eea
As pointed out in previous work \cite{gheri1998,molmer2020,biswas2020}, the coupling of the virtual cavities to the fields can be made time dependent for the purpose of creating arbitrary photon wavepackets, and therefore the Hamiltonian formulation is completely general for the photo detection process. In this way, we can calculate the evolution of the complete system with the elements of a single density matrix, instead of the multiple generalized density matrices in  eq.~(\ref{eq_gen_den}). All other discrete-continuum couplings are at position $x=0$. 
\bea
H_{b_{1(2)}-F}&=&i\sqrt{\gamma_1} \ket{F_{0(2)}}\bra{F_{1(3)}}B_{1(2)}^{\dagger}(x=0,t)+H.c.\nonumber\\
H_{F-g}&=&i\sqrt{\gamma_2} \ket{F_2}\bra{F_1}G^{\dagger}(x=0,t)+ H.c. \nonumber\\
H_{F-h}&=&i\sqrt{\gamma_4} \ket{F_4}\bra{F_3}H^{\dagger}(x=0,t)+ H.c. \nonumber\\
H_{c-d}&=&i\sqrt{\Gamma} cD^{\dagger}(x=0,t)+ H.c., 
\eea
where the field operators $G(x,t)$ and $H(x,t)$ are defined in terms of $g(\omega)$ and $h(\omega)$ 
just as the field operator $B(x,t)$ is defined in Eq.~(\ref{Bx}) in terms of $b(\omega)$.

The amplification mechanism is embodied in the parts,
\bea
\tF=\sum_{k=0,1,..,4}F_k\proj{F_k}, \label{eq_Fs} \\
H_{F-c}=i\tF(c-c^{\dagger}).\nonumber
\eea
The different eigenvalues of the operator $\tF$ drive a discrete quantum harmonic oscillator (another cavity, for example) with annihilation operator $c$ by different classical driving strengths. That driven cavity mode will contain an increasing number of excitations. We assume here $F_0 = F_1 = F_3 = 0$ so that no amplification (no driving) takes place when the molecule is in the corresponding states. The values for $F_2$ and $F_4(\neq F_2)$ are nonzero and drive the amplification process. Excitations from the driven cavity $c$ leak into the continuum mode $d(\omega)$, which can be observed ``classically'' when populated massively. Thus $d(\omega)$ contains our final ``classical'' signal. We will not analyze the macroscopic signal here and refer instead for further details to Ref.~\cite{biswas2020}, where it is shown that this type of amplification process yields minimal noise; see also \cite{epstein2021quantum}.
\subsection{Invariants of motion}
The Hamiltonian formalism preserves the basic idea of the photodetection process that is meant to be simulated. We can find some operators that commute with the Hamiltonian and are therefore conserved in time.
\bea
\mathcal{I}_{20} &=& a_1^{\dagger} a_1 + \int\limits d\omega b_1^{\dagger}(\omega) b_1(\omega) -|F_0 \rangle \langle F_0|, \nonumber\\
\mathcal{I}_{21} &=& a_1^{\dagger} a_1 + \int\limits d\omega b_1^{\dagger}(\omega) b_1(\omega) +|F_1 \rangle \langle F_1|+\int\limits d\omega g^{\dagger}(\omega) g(\omega), \nonumber\\
\mathcal{I}_{22} &=& a_2^{\dagger} a_2 + \int\limits d\omega b_2^{\dagger}(\omega) b_2(\omega) -|F_2 \rangle \langle F_2|+\int\limits d\omega g^{\dagger}(\omega) g(\omega), \nonumber\\
\mathcal{I}_{23} &=& a_2^{\dagger} a_2 + \int\limits d\omega b_2^{\dagger}(\omega) b_2(\omega) +|F_3 \rangle \langle F_3|+\int\limits d\omega h^{\dagger}(\omega) h(\omega), \nonumber\\
\mathcal{I}_{24} &=& |F_4 \rangle \langle F_4|-\int\limits d\omega h^{\dagger}(\omega) h(\omega), \nonumber
\eea
A conserved quantity of particular interest is $ \mathcal{I}_N = \frac{1}{2} \mathcal{I}_{20}+ \frac{1}{2} \mathcal{I}_{21} +\frac{1}{2} \mathcal{I}_{22}+\frac{1}{2} \mathcal{I}_{23} - \mathcal{I}_{24} + \frac{1}{2}$.
The $\frac{1}{2}$ is added here to give the invariant  $ \mathcal{I}_N$ the meaning of the number of excitations (photons). The invariant takes the values 0,1,2 for the three cases of 0,1,2 input photons, respectively.

The values of these quantities keep track of where the excitations are and whether the photons will be detected or not. For example,
an initial excitation in the $a_1$ cavity means $\mathcal{I}_{20}$ maintains a value of 0 in the entire evolution. So as the eigenvalue of $a_1^{\dagger} a_1$ decays from 1 to 0, either the eigenvalues of both $\int\limits d\omega b_1^{\dagger}(\omega) b_1(\omega)$ and $|F_0 \rangle \langle F_0|$ for $t\rightarrow\infty$ are 1 (the photon was not detected) or they are both 0 (the photon was detected). 

Similarly, $\mathcal{I}_{24}$ always equals 0, with an eigenvalue of 1 for both $|F_4 \rangle \langle F_4|$ and $\int\limits d\omega h^{\dagger}(\omega) h(\omega)$ indicating the second photon was detected, and an eigenvalue 0 indicating it was not detected (yet).
\subsection{The Liouvillian Representation}
Due to the continua in our model, the Hilbert space is infinite dimensional. However, we follow the well established practice of eliminating the continua and focus our attention on the ``system Hilbert space, $\mathcal{H}_d$'' (d=2$\times$2$\times$5=20) and are able to calculate all quantities of interest in the vector space of the linear operators, $L(\mathcal{H}_d)$ acting on the Hilbert space, $\mathcal{H}_d$. We eliminate the continua $b_1, b_2, g, h$ and obtain our Liouvillian master equation for the system density operator, $\rho_s$ comprised of discrete quantum systems $a_1, a_2, F$. Details of the exact method and validation of quantum mechanical commutation relationships can be found in the preceding paper Ref. \cite{biswas2020}. For the absorption problem, we need not include the discrete cavity mode c. The Liouvillian master equation for the chosen discrete quantum parts of the Hamiltonian is
\bea
\frac{\partial}{\partial t}\rho_s = -i[H_{sys}, \rho_s] + \mathcal{D}\left[ \rho_s  \right]. \ \ \ \ \ \ \label{eq_Lio}
\eea
Eq. \ref{eq_Lio} facilitates numerical calculation in the Schr\"odinger picture. For a collapse operator, X, the Lindblad dissipator super-operator (a map, S: $ L(\mathcal{H}_d) \rightarrow L(\mathcal{H}_d) $) acting on the system density operator, $\rho_s$ has the form, $\mathcal{D}_X \left[ \rho_s \right] = X\rho_s X^{\dagger} - \frac{1}{2} \rho_s X^{\dagger}X - \frac{1}{2}X^{\dagger}X \rho_s$. For time dependent coupling of system and environment, the collapse operator take time dependent forms \cite{molmer2019, molmer2020}. The collapse operator embodying the decay from the upper state $\ket{F_{1(3)}}$  back to the state $\ket{F_{0 (2)}}$ takes the form $X = g^*_{1(2)}(t) a_{1(2)} + \sqrt{\gamma_1} |F_{0(2)} \rangle \langle F_{1(3)}| $. A quantum jump effected by this operator indicates the corresponding photon was not detected.

The density operator in Eq. \ref{eq_Lio} can be expanded in the partial basis of the two virtual cavity populations, i.e. four basis states $\ket{n,m}$ with $n,m=0,1$ indicating the number of photons inside the cavity. This gives rise to a coupled equation set of sixteen coefficients. The complete expansion can be found in eq. \ref{eq:rho_expn_S}. With the transformation in eq. \ref{eq_trans}, we get back the equation set in eq. \ref{eq_set_ph} for the generalized density matrix operators, $\rho_{ab}(t)$ with the substitution.
\bea
u_{\alpha(\beta)}(t) = g^*_{1(2)}(t) e^{-\frac{1}{2}\int\limits_{t_0}^t dt' |g_{1(2)}(t')|^2 }, \ \ \ \ \label{eq_u2g}
\eea
which is the same result as found in \cite{molmer2019,molmer2020}. Inversion of the relationship in eq. \ref{eq_u2g} gives away the method of varying the couplings $g_{1(2)}(t)$ in time so as to generate a desired photon wavepacket $u_{\alpha(\beta)}(t)$ \cite{molmer2019, molmer2020}
\bea
g_{1(2)}(t) = \frac{u_{\alpha(\beta)}^*(t)}{\sqrt{1-\int\limits_{t_0}^t dt' |u_{\alpha(\beta)}(t')|^2 }}. \ \ \ \label{eq_g2u}
\eea

\section{Results}\label{num}

\subsection{Exponentially decaying $u_{\alpha}(t)$ and $u_{\beta}(t)$ }
\begin{figure}[h]
	\includegraphics[height=4.9cm]{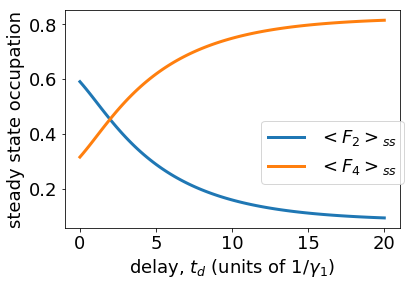}  
	\caption{\label{fig:F2 F4 delay} The steady state occupations of $F_2$ and $F_4$ level for the initial state of both cavity having one photon. Here we chose $\gamma_1=\gamma_2=\gamma_3=\gamma_4$ and $\kappa_1=\kappa_2=\gamma_1/5$. }
\end{figure}
With $u_{\alpha}(t)=\sqrt{\kappa_1} e^{-\kappa_1t/2}\Theta(t) $ and $u_{\beta}(t)=\sqrt{\kappa_2} e^{-\kappa_2t/2}\Theta(t)$, ($\Theta(t)$ being the Heaviside unit step function) either dynamical equation sets \ref{eq_set_ph} or \ref{eq_Lio} as well as quantities like absorption probabilities (eq. \ref{eq_r2424}) can be solved analytically. Through eq. \ref{eq_g2u}, we find both couplings $g_{1(2)}(t) = \sqrt{\kappa_{1(2)}}$ to be constant in time, This is the only example which can be calculated with a time-independent system Hamiltonian and collapse operators. As discussed previously, if the two photons have significant temporal overlap, the second photon may get reflected before the first is absorbed, and the molecule may end up in $|F_2\rangle$ instead of being raised all the way to $|F_4 \rangle$. If we gradually delay the second photon in time with increasing delay periods ($t_d$) and calculate the steady state populations in $\ket{F_2}$ and $\ket{F_4}$ from the Liouvillian equation each time, we find that with a longer delay the second photon is absorbed with higher probability (Fig. \ref{fig:F2 F4 delay}). The sum of the two populations is always $P^{\alpha}_{abs}(=10/11)$, the probability of the first photon being absorbed. For a delay $t_d\gg 1/\gamma_1$, the steady state occupation of the $F_4$ level reaches $P^{\alpha}_{abs} P^{\beta}_{abs}(=100/121)$, as expected.

\begin{figure}[h]
\includegraphics[height=4.9cm]{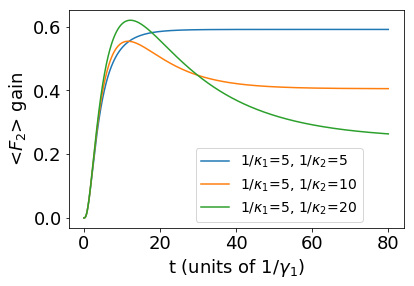} 
\caption{\label{fig:F2 occ} The occupation of the $F_2$ level as a function of time  for an initial state of both cavities having one photon.  The decay rate $\kappa_2$ of the second cavity (which determines the width in time of the second photon) is varied. All rates $\gamma$s are equal, and the rates $\kappa_{1,2}$ are given in units of $\gamma_1$. The steady-state population of $\ket{F_2}$ decreases with decreasing value of $\kappa_2$, that is, with increasing width of the second photon. The second photon moved population out of level $\ket{F_2}$.}
\end{figure}

\begin{figure}[h]
\includegraphics[height=4.9cm]{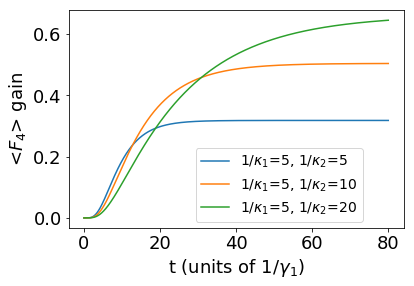} 
\caption{\label{fig:F4 occ} Same as the preceding Figure, but plotting the occupation of level $F_4$. Here the steady-state population of $\ket{F_4}$ increases with decreasing value of $\kappa_2$ since the second photon is more effective at moving population from $\ket{F_2}$ to $\ket{F_4}$.}
\end{figure}

However, the delay is not the only critical determinant of the absorption probability of the second photon. (The absorption of the first photon is completely independent of the second.) The closer to $P_{\alpha}$ the occupation in $|F_2\rangle$ has risen when the second photon arrives, the more efficient the absorption of the second photon is. So, a longer width of the second photon wavefunction would also increase the efficiency of the second photon absorption. The overlap between the two photon wave functions determines the efficiency of the second photon absorption. By making $\kappa_2$ smaller we can make the second photon wave function longer in time. In Figs. \ref{fig:F2 occ} and \ref{fig:F4 occ} respectively we plot the average occupation levels of $|F_2\rangle$ and of $|F_4\rangle$ as functions of time for different values for $\kappa_2$ that make the wavefunction of the second ($\beta$) photon longer. The same colored curves from Figs. \ref{fig:F2 occ} and \ref{fig:F4 occ} add up to $P_{\alpha}=10/11$ for $t\rightarrow\infty$.
With the larger share of the second photon coming into the detector after the first photon has already populated the $F_2$ level, the probability of a successful two-photon absorption rises. 

\subsection{Gaussian $|u_{\alpha}(t)|^2$ and $|u_{\beta}(t)|^2$ }
\begin{figure}[h]
\includegraphics[height=6.5cm]{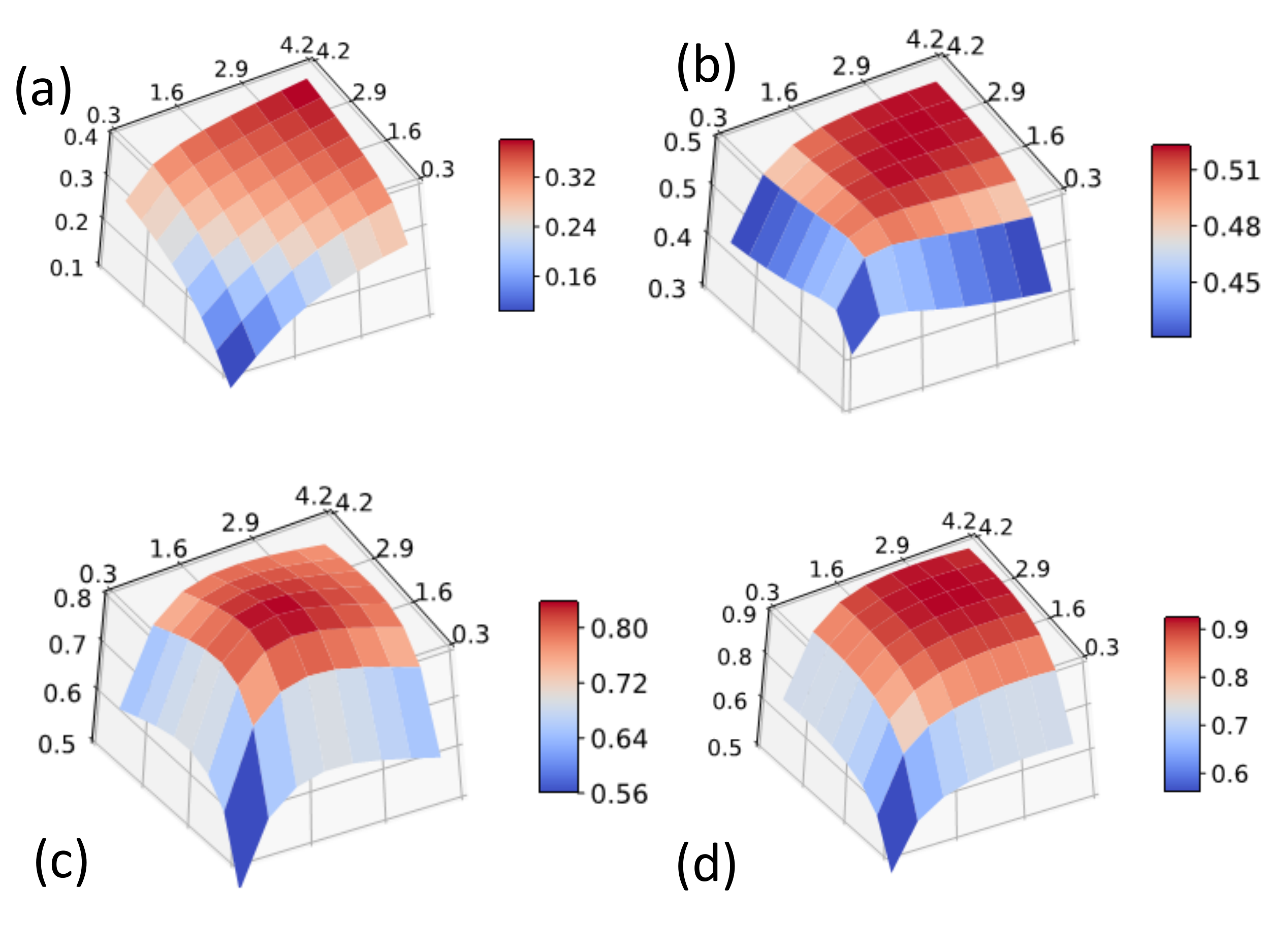} 
\caption{\label{fig:GauGau} $\mathcal{\rho}_{24,24}(\infty)$ plotted on the vertical z-axis against the standard deviations of the $\alpha$ and $\beta$ photons plotted on the two axes on the horizontal plane for (a) no time delay,  (b) $1/\gamma_1$,(c) $3/\gamma_1$, and (d) $5/\gamma_1$ delays (of the means/centres of the waveshape) of the second photon, $\beta$. }
\end{figure}

We numerically calculate the two photon absorption probability, $\mathcal{\rho}_{24,24}(\infty)$ for two real Gaussian wavefunctions with varying standard deviations and the second one delayed by different delays. The results are plotted in Fig.~\ref{fig:GauGau}. A note on the numerical method is in order here. For repeated calculations with different values of the parameters, we use eq. \ref{eq_r2424}, since it is less demanding than solving the Liouvillian equation (eq. \ref{eq_Lio}) many times. For the numerical calculation of the principal value, we use

\be
P\left(\frac{1}{\omega}\right) = \frac{1}{2}\left( \frac{1}{\omega-i\epsilon} + \frac{1}{\omega+i\epsilon} \right), \nonumber
\ee
and use a sufficiently small $\epsilon$.

 Curiously, we find that the efficiency plot is symmetric in their standard deviations for any given delay. The efficiency for a standard deviation $\sigma_1$ for photon $\alpha$ and standard deviation $\sigma_2$ for photon $\beta$ for a given time delay $t_d$ is the same as for standard deviations $\sigma_2$ for photon $\alpha$ and $\sigma_1$ photon $\beta$. The probability of the second photon absorption improves a lot as the delay is increased. For some delays there is a peak efficiency for a certain standard deviation and falls off slightly with even longer standard deviations.

\subsection{Gaussian $|u_{\alpha}(t)|^2$ and exponentially decaying $u_{\beta}(t)$ }
\begin{figure}[h]
\includegraphics[height=6.5cm]{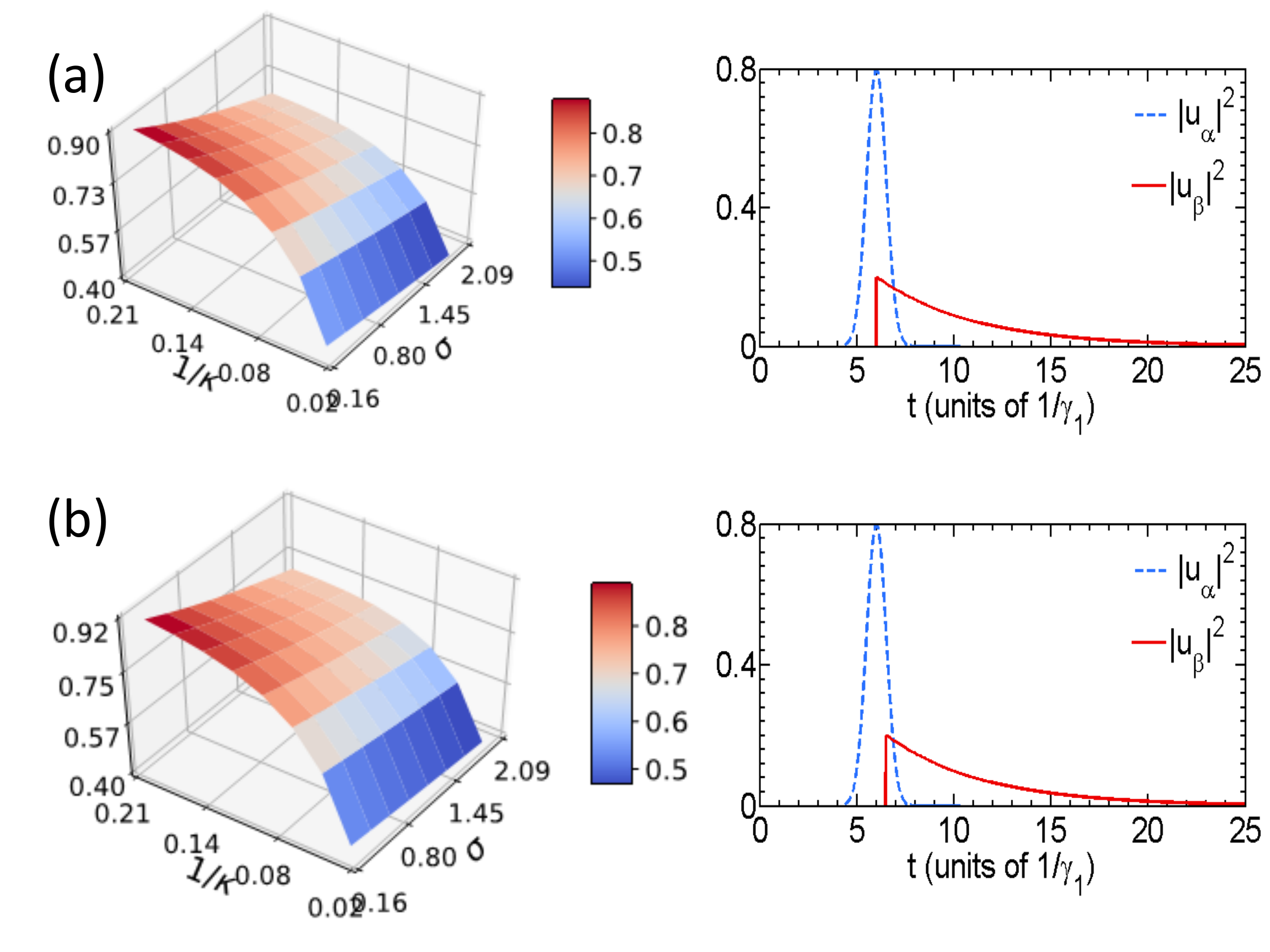} 
\caption{\label{fig:GauExp} $\mathcal{\rho}_{24,24}(\infty)$ plotted on the vertical z-axis against the standard deviations of the $\alpha$ photon Gaussian waveshape on the x axis and the inverse of the rate constant ($\kappa$) of the $\beta$ photons plotted on the y axis for (a) no time delay,  (b) $0.5/\gamma_1$ delay (delay between the mean of the Gaussian and the onset of the $\beta$ photon waveshape). }
\end{figure}

For completeness we also consider the ``mixed'' case of one Gaussian wave packet (for the first photon) and an exponentially decaying wave packet (for the second photon). The results are plotted in Fig.~\ref{fig:GauExp}
 As long as the two photon wavefunctions overlap, we get a decrease in the $\mathcal{\rho}_{24,24}(\infty)$ value with increasing the standard deviation of the $\alpha$ photon. Increase in $\kappa$ or decrease in the time constant of the $\beta$ photon increases $P^{\beta}_{\rm abs}$ and is responsible for larger $\mathcal{\rho}_{24,24}(\infty)$.

\section{Conclusions}\label{discuss}
We developed a fully quantum-mechanical model for a photon-number resolving detector that can detect up to two photons by extending the model of Ref.~\cite{biswas2020} to a five-level molecule.
Moreover, we used two different methods for treating the interaction of two photons with a quantum system---the methods developed by \cite{gheri1998} and \cite{baragiola2012n} on the one hand, and by \cite{molmer2019,molmer2020} on the other---and provided the explicit connection between the two. The former method allowed us to obtain several analytical results  in Section \ref{Synopsis} that characterize our detector, the latter method is very well suited for numerical calculations, as shown in Section \ref{num}.

The model developed in \cite{biswas2020} followed the lead by Refs.~\cite{young2018,young2018b,young2020design} in taking inspiration from visual systems appearing in biology.
It is an open question whether our current extension of that model can be found in nature as well: in particular, whether the specific {\em step-wise} two-photon absorption process we studied here occurs in the human eye, just as {\em simultaneous} two-photon absorption does occur \cite{artal2017}.

We note two extensions of our work that may be interesting.
The first  extension of our model is to another type of five-level molecule that would detect just one photon, but it would be sensitive to polarization. From the initial state we could either reach an excited state $\ket{F_{1}}$ (as in our actual model) but also an alternative excited state $\ket{F_{1'}}$ (for an orthogonally polarized photon), which would then decay to a different metastable state $\ket{F_{2'}}$. If the signal
produced in the latter state is distinguishable from that produced by $\ket{F_2}$, then this molecule would perform a polarization-sensitive single-photon measurement. It is known that some animals (insects, fish, birds) did develop polarization vision, see, e.g. Ref.~\cite{horvath2014}.
 
Second, we focused here on the case of two distinguishable input photons, with different frequencies. The case of two overlapping frequencies (relevant when the two molecular transitions would have nearly equal transition frequencies) would reveal two additional features. Both input photons would be able to drive the two transitions, and the final expression for the two-photon absorption amplitude would contain two terms, corresponding to two different time orders in which the ``first'' and ``second'' photon could be absorbed. Those two terms may interfere destructively. That type of effect is certainly interesting but known \cite{schrama1991,vanenk2017b}. Moreover, the POVM element would involve projections onto {\em entangled} two-photon input states, like it does for standard two-photon absorption \cite{fei1997}.

\section*{Acknowledgments}
This work is supported by funding from
DARPA under
Contract No. W911NF-17-1-0267.

\appendix
\section{Emulating Photon wavepackets with Auxiliary Cavities}

We outline the systematic process of deriving the transformations between the generalized density matrix operators and coefficients (in the expansion in eq. \ref{eq:rho_expn_S}) in the Hamiltonian formulation. Unlike Gheri et al, we do not introduce a detuning of the auxiliary cavities for the emulation of the generalized density matrix equations (Refs. \cite{molmer2019,molmer2020} did not either). Gheri et al addressed the mapping for the problem of photons (one or few) in a single continuum. For the problem of two photons residing in two continua (or even more complex scenarios), the procedure outlined here can find the mapping between the two formalisms (generalized density matrix operator and Hamiltonian formulation) systematically. The generalized density matrix equations found for the system described in sections III and IV are

\bea
\dot{\mathcal{\rho}}_{2,2}(t)= \mathcal{L}\{ \mathcal{\rho}_{2,2}\}(t) + \left[ \sqrt{\gamma_3} u_{\beta}(t) [ \mathcal{\rho}_{\alpha 2}(t), |F_3\rangle \langle F_2|(t_0)] \right. \ \ \ \ \nonumber\\
\left. +\sqrt{\gamma_1} u_{\alpha}(t) [ \mathcal{\rho}_{\beta 2}(t), |F_1\rangle \langle F_0|(t_0)] + H.c. \right] \ \ \ \ \ \ \ \ \ \ \ \ \ \ \ \ \ \ \ \ \ \ \ \ \ \nonumber\\
\dot{\mathcal{\rho}}_{\alpha,2}(t) = \mathcal{L}\{ \mathcal{\rho}_{\alpha,2}\}(t) - \sqrt{\gamma_3} u^*_{\beta}(t) [ \mathcal{\rho}_{\alpha,\alpha}, |F_2 \rangle \langle F_3|(t_0) ] + \ \ \ \ \ \nonumber\\
\sqrt{\gamma_1} u_{\alpha}(t) [\mathcal{\rho}_{0,2}, |F_1\rangle \langle F_0|(t_0)]- \sqrt{\gamma_1} u^*_{\alpha}(t) [ \mathcal{\rho}_{\alpha,\beta}, |F_0 \rangle \langle F_1|(t_0)] \ \nonumber\\
\dot{\mathcal{\rho}}_{\beta,2}(t) = \mathcal{L}\{ \mathcal{\rho}_{\beta,2}\}(t) - \sqrt{\gamma_1} u^*_{\alpha}(t) [ \mathcal{\rho}_{\beta,\beta}, |F_0 \rangle \langle F_1|(t_0) ] + \ \ \ \ \ \nonumber\\
\sqrt{\gamma_3} u_{\beta}(t) [\mathcal{\rho}_{0,2}, |F_3\rangle \langle F_2|(t_0)]- \sqrt{\gamma_3} u^*_{\beta}(t) [ \mathcal{\rho}_{\beta,\alpha}, |F_2 \rangle \langle F_3|(t_0)] \nonumber\\
\dot{\mathcal{\rho}}_{\alpha, \alpha}(t)= \mathcal{L}\{ \mathcal{\rho}_{\alpha, \alpha}\}(t) -\sqrt{\gamma_1} u_{\alpha}(t) \left[ |F_1 \rangle \langle F_0 |(t_0), \mathcal{\rho}_{0, \alpha}(t)  \right] \ \ \ \ \nonumber\\
+ \sqrt{\gamma_1} u^*_{\alpha}(t) \left[ |F_0 \rangle \langle F_1 |(t_0), \mathcal{\rho}_{\alpha, 0}(t)  \right] \ \ \ \ \nonumber\\
\dot{\mathcal{\rho}}_{\beta,\beta}(t)= \mathcal{L}\{ \mathcal{\rho}_{\beta,\beta}\}(t) - \sqrt{\gamma_3} u_{\beta}(t) \left[ |F_3 \rangle \langle F_2 |(t_0), \mathcal{\rho}_{0,\beta}(t)  \right] \ \ \ \ \nonumber\\
+\sqrt{\gamma_3} u^*_{\beta}(t) \left[ |F_2 \rangle \langle F_3 |(t_0), \mathcal{\rho}_{\beta, 0}(t)  \right] \ \ \ \ \nonumber\\
\dot{\mathcal{\rho}}_{\alpha, \beta}(t)= \mathcal{L}\{ \mathcal{\rho}_{\alpha, \beta}\}(t) -\sqrt{\gamma_1} u_{\alpha}(t) \left[ |F_1 \rangle \langle F_0 |(t_0), \mathcal{\rho}_{0, \beta}(t)  \right] \ \ \ \ \nonumber\\
+ \sqrt{\gamma_3} u^*_{\beta}(t) \left[ |F_2 \rangle \langle F_3 |(t_0), \mathcal{\rho}_{\alpha, 0}(t)  \right] \ \ \ \ \nonumber\\
\dot{\mathcal{\rho}}_{0,2}(t) = \mathcal{L}\{ \mathcal{\rho}_{0,2}\}(t)  \ \ \ \ \ \ \ \ \ \ \ \ \ \ \ \ \ \ \ \ \ \ \ \ \ \ \ \ \ \ \ \ \ \ \ \ \ \ \ \ \ \ \ \ \ \ \ \ \ \nonumber\\
-\sqrt{\gamma_3} u^*_{\beta}(t) [\mathcal{\rho}_{0,\alpha}, |F_2 \rangle \langle F_3 |(t_0)] - \sqrt{\gamma_1} u^*_{\alpha}(t)  [\mathcal{\rho}_{0,\beta}, |F_0 \rangle \langle F_1| (t_0)] \nonumber\\
\dot{\mathcal{\rho}}_{0,\alpha}(t) = \mathcal{L}\{ \mathcal{\rho}_{0,\alpha}\}(t) + \sqrt{\gamma_1} u^*_{\alpha}(t) [ \mathcal{\rho}_{0,0}, |F_0 \rangle \langle F_1|(t_0)] \ \ \ \ \ \ \ \ \ \nonumber\\
\dot{\mathcal{\rho}}_{0,\beta}(t) = \mathcal{L}\{ \mathcal{\rho}_{0,\beta}\}(t) + \sqrt{\gamma_3} u^*_{\beta}(t) [ \mathcal{\rho}_{0,0}, |F_2 \rangle \langle F_3|(t_0)] \ \ \ \ \ \ \ \ \ \nonumber\\
\dot{\mathcal{\rho}}_{0,0}(t) = \mathcal{L}\{ \mathcal{\rho}_{0,0}\}(t) \ \ \ \ \ \ \ \ \ \ \ \ \ \ \ \ \ \ \ \ \ \ \ \ \ \ \ \ \ \ \ \ \ \ \ \ \ \ \ \ \ \ \ \ \ \ \ \ \label{eq_set_ph}
\eea

$\mathcal{D}$ is the Lindblad dissipator superoperator and its explicit form depends on the number and nature of the baths coupled to the system. Due to the coupling of the total of four continua, we get four collapse operators for the Lindblad dissipator superoperator, $\mathcal{D}$, namely $\sqrt{\gamma_1} |F_0 \rangle \langle F_1|$, $\sqrt{\gamma_2} |F_2 \rangle \langle F_1|$, $\sqrt{\gamma_3} |F_2 \rangle \langle F_3|$, and $\sqrt{\gamma_4} |F_4 \rangle \langle F_3|$. The density matrix in the Hamiltonian formulation can be expanded in the complete basis (time dependence restricted to the expansion coefficients):


\bea
\mathcal{\rho}(t) =  \ \ \ \ \ \ \ \ \ \ \ \ \ \  \ \ \ \ \ \ \ \ \ \ \ \ \ \  \ \ \ \ \ \ \ \ \ \ \ \ \ \  \ \ \ \ \ \ \ \ \ \ \ \ \ \  \ \ \ \ \ \  \nonumber\\
 \tilde{\mathcal{\rho}}_{2,2}(t)| 0,0 \rangle \langle 0,0 | + \tilde{\mathcal{\rho}}_{\alpha, \alpha}(t)| 0,1 \rangle \langle 0,1 |+  \tilde{\mathcal{\rho}}_{\beta,\beta}(t) | 1,0 \rangle \langle 1,0 | \nonumber\\
+ \tilde{\mathcal{\rho}}_{\beta,2}(t) | 1,0 \rangle \langle 0,0 |+ \tilde{\mathcal{\rho}}_{0,\alpha}(t) | 1,1 \rangle \langle 0,1 | \nonumber\\
+  \tilde{\mathcal{\rho}}_{2,\beta}(t) | 0,0 \rangle \langle 1,0 | + \tilde{\mathcal{\rho}}_{\alpha, 0}(t)| 0,1 \rangle \langle 1,1 | \nonumber\\
+ \tilde{\mathcal{\rho}}_{\alpha,2}(t) | 0,1 \rangle \langle 0,0 |+ \tilde{\mathcal{\rho}}_{0,\beta}(t) | 1,1 \rangle \langle 1,0 |  \nonumber\\
+ \tilde{\mathcal{\rho}}_{2,\alpha}(t)| 0,0 \rangle \langle 0,1 |+ \tilde{\mathcal{\rho}}_{\beta,0}(t)| 1,0 \rangle \langle 1,1 | \nonumber\\
+\tilde{\mathcal{\rho}}_{0,0}(t)| 1,1 \rangle \langle 1,1 | + \tilde{\mathcal{\rho}}_{0,2}(t)| 1,1 \rangle \langle 0,0 |+ \tilde{\mathcal{\rho}}_{2,0}(t) | 0,0 \rangle \langle 1,1 | \nonumber\\
+ \tilde{\mathcal{\rho}}_{\beta,\alpha}(t) | 1,0 \rangle \langle 0,1 | + \tilde{\mathcal{\rho}}_{\alpha,\beta}(t) | 0,1 \rangle \langle 1,0 | \ \ \ \ \ \ \ \ \ \ \ \ \ \ \label{eq:rho_expn_S} 
\eea
Here a state $\ket{n,m}$ for $n,m\in \{0,1\}$ indicates the number of photons in the two cavities, respectively. Since we start the cavities with one photon each, an input photon in mode $\alpha$ or $\beta$ will correspond to a cavity state $\ket{n,m}$ with $n=0$ or $m=0$ respectively (the photons have leaked out of their cavities). The $F$-operators in the Hamiltonian and Lindbladian of eq. \ref{eq_Lio} act on the expansion coefficients, $\tilde{\mathcal{\rho}}_{i,j}(t)$ and the cavity mode annihilation $a_1,a_2$ operators act on the basis elements $|n,m\rangle$. Using the expansion, eq. \ref{eq:rho_expn_S} in eq. \ref{eq_Lio}, we get a set of coupled differential equations for the 16 coefficients. For example, the two coefficients, $\tilde{\mathcal{\rho}}_{\beta, 2}(t)$ and $\tilde{\mathcal{\rho}}_{0, \alpha}$ have the coupled equations,
\bea
\dot{\tilde{\mathcal{\rho}}}_{\beta, 2}(t) = \left( \mathcal{D}_{ \sqrt{\gamma_1} |F_0 \rangle \langle F_1|} + \mathcal{D}_{\sqrt{\gamma_2} |F_2 \rangle \langle F_1|} + \mathcal{D}_{\sqrt{\gamma_3} |F_2 \rangle \langle F_3|} \right. \nonumber\\
\left. + \mathcal{D}_{\sqrt{\gamma_4} |F_4 \rangle \langle F_3|} \right) \tilde{\mathcal{\rho}}_{\beta, 2}(t) - \frac{1}{2}|g_2(t)|^2 \left[ \tilde{\mathcal{\rho}}_{\beta, 2} \right] + |g_1(t)|^2 \tilde{\mathcal{\rho}}_{0, \alpha}(t)  \nonumber\\
+ g_1(t)\sqrt{ \gamma_1} \left[ |F_0 \rangle \langle F_1|, \tilde{\mathcal{\rho}}_{\beta, \beta}(t) \right] -g_2(t) \sqrt{\gamma_3} \left[ \tilde{\mathcal{\rho}}_{\beta, \alpha} , |F_2 \rangle \langle F_3|\right] \nonumber\\
-g_2^*(t)\sqrt{ \gamma_3} \left[  |F_3 \rangle \langle F_2|, \tilde{\mathcal{\rho}}_{0, 2}(t) \right]   \ \ \ \ \ \ \ \ \ \ \ \label{eq_rb2}\\
\dot{\tilde{\mathcal{\rho}}}_{0, \alpha}(t) = \left(  \mathcal{D}_{\sqrt{\gamma_1} |F_0 \rangle \langle F_1|}+\mathcal{D}_{\sqrt{\gamma_3}|F_2 \rangle \langle F_3| } \right. \nonumber\\
\left. - |g_1(t)|^2 - \frac{1}{2}|g_2(t)|^2 \right) \tilde{\mathcal{\rho}}_{0, \alpha}  \nonumber\\
+ g_1(t)\sqrt{ \gamma_1} \left[ |F_0 \rangle \langle F_1|, \tilde{\mathcal{\rho}}_{0, 0}(t) \right] \ \ \ \ \ \ \ \ \ \ \ \ \ \ \ \ \ \ \ \ \ \ \ \label{eq_r0a}
\eea

 The following transformation from the tilde operators gives us back the set of differential equations in eq.  \ref{eq_set_ph} with the generalized density matrix operators defined previously in eq. \ref{eq_gen_den}. This comes with the substitutions in eq. \ref{eq_g2u}. 

\bea
\mathcal{\rho}_{0, 0}(t) &=& \tilde{\mathcal{\rho}}_{0, 0}(t) e^{\int dt (|g_1(t)|^2+|g_2(t)|^2)} \nonumber\\
\mathcal{\rho}_{\alpha, \alpha}(t) &=& \left[ \tilde{\mathcal{\rho}}_{\alpha, \alpha}(t) +\tilde{\mathcal{\rho}}_{0, 0}(t) \right] e^{\int dt |g_1(t)|^2}  \nonumber\\
\mathcal{\rho}_{\beta, \beta}(t) &=& \left[ \tilde{\mathcal{\rho}}_{\beta, \beta}(t)+ \tilde{\mathcal{\rho}}_{0, 0}(t) \right] e^{\int dt |g_2(t)|^2}  \nonumber\\
\mathcal{\rho}_{2,2} &=& \left[ \tilde{\mathcal{\rho}}_{2,2} + \tilde{\mathcal{\rho}}_{\alpha, \alpha}(t) + \tilde{\mathcal{\rho}}_{\beta, \beta}(t) \right] \nonumber\\
\mathcal{\rho}_{0, \alpha}(t) &=& \tilde{\mathcal{\rho}}_{0, \alpha}(t) e^{\int dt (|g_1(t)|^2+\frac{1}{2}|g_2(t)|^2)}  \nonumber\\
\mathcal{\rho}_{\beta, 2}(t) &=& \left[ \tilde{\mathcal{\rho}}_{\beta, 2}(t) + \tilde{\mathcal{\rho}}_{0, \alpha}(t) \right] e^{\int dt \frac{1}{2}|g_2(t)|^2}  \nonumber\\
\mathcal{\rho}_{0, \beta}(t) &=& \tilde{\mathcal{\rho}}_{0, \beta}(t) e^{\int dt (\frac{1}{2}|g_1(t)|^2+|g_2(t)|^2)}  \nonumber\\
\mathcal{\rho}_{\alpha, 2}(t) &=& \left[ \tilde{\mathcal{\rho}}_{\alpha, 2}(t)+\tilde{\mathcal{\rho}}_{0, \beta}(t) \right] e^{\int dt \frac{1}{2}|g_1(t)|^2 }  \nonumber\\
\mathcal{\rho}_{\alpha, \beta} &=& \tilde{\mathcal{\rho}}_{\alpha, \beta} e^{\int dt \frac{1}{2}(|g_1(t)|^2+|g_2(t)|^2)} \nonumber\\
\mathcal{\rho}_{0, 2}(t) &=& \tilde{\mathcal{\rho}}_{0, 2}(t) e^{\int dt \frac{1}{2}(|g_1(t)|^2+|g_2(t)|^2)}  \ \ \ \ \ \ \ \ \label{eq_trans}
\eea

The coefficients in eq. \ref{eq:rho_expn_S} can actually be given the meaning of a density matrix element with their usual meaning. The Hamiltonian formulation helps us write the density operator with trace of 1 that embodies all the generalized density operators in eq. \ref{eq_set_ph} (the diagonal(off-diagonal) ones each have a preserved trace of 1(0) each).

\bibliography{two_photon6}
\end{document}